\documentclass[11pt]{article}

\usepackage{amssymb}
\usepackage{amsmath}
\usepackage[dvips]{epsfig}
\usepackage{caption2}
\usepackage{graphicx}
\usepackage[colorlinks=true,linkcolor=blue,citecolor=red]{hyperref}

\newtheorem{Lemma}{Lemma}[section]
\newtheorem{Theorem}{Theorem}
\newtheorem{Proposition}[Lemma]{Proposition}
\newtheorem{Corollary}[Lemma]{Corollary}
\newtheorem{Remark}[Lemma]{Remark}

\newtheorem{Hypothesis}[Lemma]{Hypothesis}

\newenvironment{Proof}%
 {\begin{trivlist} \item[]{\bf Proof. }}%
 {\hspace*{\fill}$\rule{.4\baselineskip}{.4\baselineskip}$\end{trivlist}}

 {\begin{trivlist}\item[]\textbf{Acknowledgments }}{\end{trivlist}}

\makeatletter\@addtoreset{figure}{section}\makeatother

\makeatletter \@addtoreset{equation}{section} \makeatother

\newcommand{\R}{\mathbb{R}}
\newcommand{\C}{\mathbb{C}}
\newcommand{\Q}{\mathbb{Q}}
\newcommand{\N}{\mathbb{N}}
\newcommand{\Z}{\mathbb{Z}}

\def\Re{\mathop{\mathrm{Re}}}

\renewcommand{\leq}{\leqslant}
\renewcommand{\geq}{\geqslant}

\newfam\bifam
\font\tenbi=cmmib10 scaled \magstep1 \font\sevenbi=cmmib10 at 11pt
\font\fivebi=cmmib10 at 6pt \textfont\bifam = \tenbi
\scriptfont\bifam = \sevenbi \scriptscriptfont\bifam= \fivebi

\begin{document}

\thispagestyle{empty}

\title{\bf  Counting unstable eigenvalues in Hamiltonian spectral problems via commuting operators}

\author{\Large Mariana Haragus$^{1}$, Jin Li$^{2}$, and Dmitry E. Pelinovsky$^{2}$ \\[0.5ex]
{\small $^{1}$ Institut FEMTO-ST \& LMB, Univ. Bourgogne Franche--Comt\'e, 25030, Besan\c con, France } \\
{\small $^{2}$ Department of Mathematics, McMaster University, Hamilton, Ontario, L8S 4K1, Canada} }
\date{}

\maketitle

\begin{abstract}
We present a general counting result for the unstable eigenvalues of linear operators of the form $JL$ in which $J$ and $L$ are skew- and self-adjoint operators, respectively. Assuming that there exists a self-adjoint operator $K$ such that the operators $JL$ and $JK$ commute, we prove that the number of unstable eigenvalues of $JL$ is bounded by the number of nonpositive eigenvalues of~$K$. As an application, we discuss the transverse stability of one-dimensional periodic traveling waves in the classical KP-II (Kadomtsev--Petviashvili) equation. We show that these one-dimensional periodic waves are transversely spectrally stable with respect to general two-dimensional bounded perturbations, including periodic and localized perturbations in either the longitudinal or the transverse direction, and that they are transversely linearly stable with respect to doubly periodic perturbations.
\end{abstract}

\section{Introduction}

Linearized operators arising in stability studies for Hamiltonian systems have a typical product structure $JL$ in which $J$ is a skew-adjoint operator and $L$ a self-adjoint operator. Well-known results show that, under suitable conditions, the number of unstable eigenvalues (i.e., the eigenvalues with positive real part) of the operator $JL$ is bounded by the number of nonpositive eigenvalues of the self-adjoint operator $L$ (e.g., see \cite{CP, HK, KKS} and the references therein). In particular, if the operator $L$ is positive-definite this immediately implies that $JL$ has no unstable spectrum.
Since typically $L$ is related to the Hessian operator of an energy functional that is conserved in the time evolution of the Hamiltonian system,
besides spectral stability, one can also conclude on nonlinear, orbital stability. Such results have been extensively used in the analysis of the stability of nonlinear waves (e.g., see the books \cite{AP, KaP}).

While these arguments work very well for solitary waves, for periodic waves they allow, so far, to only understand stability with respect to co-periodic perturbations (i.e., which have the same period as that of the wave). The main difficulty in the case of periodic waves, is the fact that the number of negative eigenvalues of the operator $L$ increases when the period of the perturbations is an increasing multiple of the period of the wave, and that $L$ has negative essential spectrum when the perturbations are localized. These are serious obstacles in controlling unstable eigenvalues and then proving stability of periodic waves for arbitrary bounded perturbations.

In this paper we generalize this classical eigenvalue counting result by showing that the operator $L$ can be replaced by another self-adjoint operator $K$, provided the operators $JL$ and $JK$ commute. More precisely, under suitable assumptions, we prove that the number of unstable eigenvalues of the operator $JL$ is bounded by the number of nonpositive eigenvalues of the self-adjoint operator $K$. In applications, and in particular for periodic waves, when the operator $L$ has too many negative eigenvalues to conclude on stability, one could then try to construct such an operator $K$ with less negative spectrum.

Very recently, the idea of using a positive definite operator $K$ has been exploited
in \cite{b1} and \cite{Decon1,gp} and allowed the authors to show
the orbital stability of periodic waves with respect to
subharmonic perturbations (i.e., the period of the perturbations is an integer multiple of the period of the wave)
for the Korteweg-de Vries (KdV) and the cubic nonlinear Schr\"odinger (NLS) equations, respectively.
In these works, the construction of $K$ was strongly related to the integrability properties of these equations, and more precisely to the existence of a higher-order conserved quantity whose Hessian provided the positive definite operator~$K$. In general,
finding such an operator $K$ for a nonintegrable equation is a  nontrivial task.

As an application of the general result, we discuss the transverse (spectral and linear) stability of one-dimensional periodic
traveling waves in a model equation derived by Kadomtsev and Petviashvili in 1970 \cite{KP}.
Thanks to the scaling properties
of this model equation, we may take it in the following normalized form
\begin{equation}
\label{KP}
(u_t + 6 u u_x + u_{xxx})_x + u_{yy} = 0,
\end{equation}
where the subscripts denote partial derivatives with respect to the spatial variables
$(x,y)$ and the temporal variable $t$. This equation is referred to as the KP-II equation, where
the index II stands for the version relevant to the case of negative transverse dispersion. The KP-I equation is obtained by replacing the positive sign in front of the term $u_{yy}$ by a negative sign, and it is relevant to the case of positive transverse dispersion. Both versions of the KP equation are two-dimensional extensions of the KdV equation
\begin{equation}
\label{kdv}
u_t + 6 u u_x + u_{xxx} = 0,
\end{equation}
that governs one-dimensional nonlinear waves in the longitudinal direction of the $x$ axis. Just like the KdV equation, the KP-II and KP-I equations arise as particular models in the classical water-wave problem, in the cases of small and large surface tension, respectively.

The KP equations quickly became very popular due to their integrability
properties \cite{ZS}, including a rich family of exact solutions, a bi-Hamiltonian structure  and the recursion
operator, a countable set of conserved quantities and symmetries, as well as the inverse scattering transform
techniques. At the same time, they became popular in the analysis of the stability of nonlinear waves, both relying upon functional-analytic methods and integrability techniques. As a model equation for surface water waves, some of the obtained results were  extended
to the Euler equations describing the full hydrodynamic problem \cite{Groves,mh,rt}.

Stability properties of traveling waves are quite different for the two versions of the KP equation. While both periodic and solitary waves are transversely unstable in the KP-I equation (e.g., see recent works \cite{Hakkaev,Matt,RT2,RT3} and the references therein), it is expected that they are transversely stable in the KP-II equation \cite{AS,KP}. Numerical evidences of these stability properties can be found for instance in \cite{KS,Klein}. For the case of solitary waves, the transverse nonlinear stability has been recently proved for periodic transverse perturbations in \cite{Mizumachi1},
 and for fully localized perturbations in \cite{Mizumachi2}. In contrast, there are few analytical results for periodic waves for which, in particular, the question of transverse nonlinear stability is open.

By using a linearized version of the dressing method from \cite{ZS}, explicit eigenfunctions
of the spectral stability problem associated with the periodic waves of the KP-II equation (\ref{KP}) were constructed in \cite{KSF}.
Completeness of the eigenfunctions and generalizations to the case of oblique transverse perturbations
were elaborated few year later \cite{Spector}. The results obtained by this method rely on explicit computations
involving Jacobi elliptic functions for the periodic waves and the associated Jost functions,
which are hard to check or confirm.
An alternative approach, based on the classical counting result for the unstable eigenvalues of linear operators of the form $JL$ mentioned above, has been recently discussed in \cite{Haragus}. It turns out that in the case of the periodic waves of the KP-II equation the self-adjoint operator $L$ has unbounded spectrum for both below and above. Consequently, this eigenvalue counting only allows to obtain a partial result, showing spectral stability of small-amplitude periodic waves with respect
to perturbations which are co-periodic in the direction of propagation $x$, and have long wavelengths in the transverse direction $y$.

In the present work, we show that the general counting result in which the operator $L$ is replaced by a suitably chosen operator $K$ allows us to give a complete proof of transverse spectral stability of periodic waves for general two-dimensional bounded perturbations. As a consequence, we also show that these periodic waves are transversely linearly stable with respect to doubly periodic perturbations, which are subharmonic and have zero mean in the direction of propagation $x$ and have an arbitrary, but fixed, period in the transverse direction $y$.
The main challenge of our method is the construction of a self-adjoint operator $K$ such that the operators $JL$ and $JK$ commute and which has a minimum number of negative eigenvalues. The best situation arises when the operator $K$ is positive, this property implying directly transverse stability.

One way of finding a self-adjoint operator $K$ satisfying the commutativity property is with the help of the
conserved quantities of the KP-II equation, as this has been done for the KdV and NLS equations in \cite{Decon1,b1,gp}.
The self-adjoint operator $L$ is related to the Hessian operator of the standard energy functional expanded
at the periodic traveling wave. Similarly, the self-adjoint
operator $K$ can be found from the Hessian operator of a higher-order energy functional,
as for instance the one used in the proof of global well-posedness for the KP-I equation \cite{Z1,Z2}.
Then the operators $JL$ and $JK$ commute.

For the KdV and NLS equations, neither $L$ and $K$ are positive operators,
but a suitable linear combination of these operators is positive \cite{Decon1,b1,gp}.
We found rather surprising that this is not the case for the KP-II equation,
when $K$ is constructed from a higher-order energy functional. In order to avoid
this obstacle, we start with the operator $K$ obtained for the KdV equation
and find an operator $K$ for the KP-II equation by a direct search from the commutativity relation.
Then we show that a suitable linear combination of $L$ and $K$ is indeed a positive operator.
However, this self-adjoint operator $K$ constructed directly from the commutativity relation does not seem to be related to the Hessian operator of some higher-order conserved quantity of the KP-II equation.
In particular, we cannot use this construction to also conclude on the nonlinear, orbital stability of these periodic waves, which remains an open problem.

The paper is organized as follows. We present the general counting result for unstable eigenvalues in Section~2.
In Section~3, we discuss the transverse spectral and linear stability problems for the periodic waves of the KP-II equation
and state the main results. The proofs of these results are given in Section~4. We conclude with a discussion of
the transverse nonlinear stability problem in Section~5.

\noindent
{\bf Acknowledgements.} M. Haragus was partially supported by the ANR project BoND (ANR-13-BS01-0009-01).
J. Li was supported by the MSc scholarship at McMaster University. D. Pelinovsky was supported by the NSERC
Discovery grant.

\section{Abstract counting result}

Here we present the general counting result for the unstable eigenvalues of
an operator $JL$ with $J$ and $L$ being skew- and self-adjoint operators, respectively.

Following a standard terminology, for a linear operator $A$, we denote by $\sigma_s(A)$, $\sigma_c(A)$, and $\sigma_u(A)$, the subsets of the spectrum $\sigma(A)$ of $A$ lying in the open left-half complex plane, on the imaginary axis, and in the open right-half complex plane, respectively.
More precisely, we denote
\begin{eqnarray*}
\sigma_s(A) & = & \{\lambda\in\sigma(A)\;;\; \Re\lambda<0\},\\
\sigma_c(A) & = & \{\lambda\in\sigma(A)\;;\; \Re\lambda=0\},\\
\sigma_u(A) & = & \{\lambda\in\sigma(A)\;;\; \Re\lambda>0\},
\end{eqnarray*}
and refer to these sets as the stable, central, and unstable spectra of $A$, respectively. Further, we  denote by $\mathrm n_s(A)$, $\mathrm n_c(A)$, and $\mathrm n_u(A)$, the dimension of the spectral subspaces associated to $\sigma_s(A)$, $\sigma_c(A)$, and $\sigma_u(A)$, respectively, if these exist. Recall that in the case of a finite spectral subset consisting only of isolated eigenvalues with finite algebraic multiplicities, the corresponding spectral subspace is finite-dimensional, and its dimension is given by the number of eigenvalues counted with algebraic multiplicities.

\begin{Hypothesis}\label{h:1}
Consider a Hilbert space $\mathcal H$  equipped with a scalar product $\langle\cdot,\cdot\rangle$.
Assume that $J$, $L$, and $K$ are closed linear operators acting in $\mathcal H$ with the following properties.
  \begin{enumerate}
  \item $J$ is a skew-adjoint operator ($J^*=-J$) with bounded inverse.
  \item $L$ and $K$ are self-adjoint operators ($L^*=L$ and $K^*=K$) such that the operators $JL$ and $JK$ commute, i.e., the operators $(JL)(JK)$ and $(JK)(JL)$ have the same domain $\mathcal D\subset \mathcal H$, and
    \begin{equation}\label{e:jljk}
(JL)(JK)u = (JK)(JL)u, \quad \forall\ u\in \mathcal D.
    \end{equation}
  \item The nonpositive spectrum $\sigma_s(K)\cup\sigma_c(K)$ of the self-adjoint operator $K$ consists, at most, of a finite number of isolated eigenvalues with finite  multiplicities.
  \item The unstable spectrum $\sigma_u(JL)$ of the operator $JL$ consists, at most, of isolated eigenvalues with finite algebraic multiplicities, and the generalized eigenvectors associated to these eigenvalues belong to the domain of the operator $JK$.
  \end{enumerate}
\end{Hypothesis}

A well-known property of the spectrum of the operator $JL$ is that it is symmetric with respect to the imaginary axis because  $J$ and $L$ are skew- and self-adjoint operators, respectively (e.g., see \cite[Proposition 2.5]{HK}). In particular, eigenvalues of $JL$ lying outside the imaginary axis arise in pairs of eigenvalues ($\lambda,-\overline\lambda$) with the same algebraic multiplicity, so that we have a one-to-one correspondence between the spectral subsets  $\sigma_s(JL)$ and  $\sigma_u(JL)$.

\begin{Remark}
  \label{r:1}
\begin{enumerate}
\item The invertibility of the operator $J$ implies that we can replace the equality \eqref{e:jljk} by the equivalent equality
  \[
(LJK)u = (KJL)u, \quad \forall\ u\in \mathcal D.
  \]
\item In the case of differential operators, as the ones which will be considered in the next section, the second part of the Hypothesis~\ref{h:1}~(iv) can be easily checked using the property that generalized eigenvectors of differential equations are often smooth functions. Alternatively, we can replace this hypothesis by slightly stronger hypotheses on the domain of the operator $JK$, as for instance that the domain of the operator $(JL)^n$ is included in the domain of $JK$, for some positive integer~$n$. Clearly, this property implies that the generalized eigenvectors of $JL$ belong to the domain of $JK$.
\end{enumerate}
\end{Remark}

The key step in the proof of our main result is the following property which holds for isolated eigenvalues of the operator $JL$ under the assumptions (i) and (ii) of Hypothesis~\ref{h:1}, only.

\begin{Lemma}\label{l:1}
  Under the assumptions (i) and (ii) of Hypothesis~\ref{h:1}, if $\lambda$ and $\sigma$ are isolated eigenvalues of $JL$ with finite algebraic multiplicities and if
  \begin{enumerate}
  \item $\lambda+\overline\sigma \not=0$,
  \item the spectral subspaces $E_\lambda$ and $E_\sigma$ associated to the the eigenvalues $\lambda$ and $\sigma$, respectively, are contained in the domain of the operator $JK$,
  \end{enumerate}
  then
  \begin{equation}\label{e:1}
\langle Ku,v\rangle = 0, \quad \forall\  u\in E_\lambda, \  v\in E_\sigma.
\end{equation}
\end{Lemma}

\begin{Proof}
The eigenvalues  $\lambda$ and $\sigma$ are isolated and have finite multiplicities, so that there exist finite bases of the associated spectral spaces $E_\lambda$ and $E_\sigma$, which consist of chains of generalized eigenvectors $\{u_1,\dots,u_n\}$ and $\{v_1,\dots,v_m\}$, respectively, satisfying
    \begin{eqnarray*}
JLu_i=\lambda u_i + u_{i-1},\  u_0=0,\ i=1,\dots,n,\\
JLv_j=\sigma v_j + v_{j-1},\  v_0=0,\ j=1,\dots,m.
      \end{eqnarray*}
It is  sufficient to prove \eqref{e:1} for $u=u_i$, $v=v_j$,
$i=1,\dots,n$, $j=1,\dots,m$. We will proceed by induction upon $i$ and $j$.

Using successively the fact that $u_i$ belong to the domain of $JK$, the commutativity of $JL$ and $JK$, and the invertibility of $J$ we obtain
  \begin{eqnarray*}
JLu_i = \lambda u_i +u_{i-1} & \Rightarrow &
  JKJLu_i=\lambda JKu_i + JKu_{i-1} \\
  & \Rightarrow &   JLJKu_i=\lambda JKu_i + JKu_{i-1} \\ & \Rightarrow &
  LJKu_i=\lambda Ku_i + Ku_{i-1}.
  \end{eqnarray*}
The last equality implies that
  \[
  \lambda\langle Ku_i,v_j\rangle = \langle LJKu_i,v_j\rangle -
  \langle Ku_{i-1},v_j\rangle,
  \]
and since $L$ and $K$ are self-adjoint operators and $J$ is a skew-adjoint operator, we also have the equality
  \[
  \overline\sigma\langle Ku_i,v_j\rangle =
\langle Ku_i, JLv_j\rangle - \langle Ku_i, v_{j-1}\rangle =
- \langle LJKu_i,v_j\rangle - \langle Ku_i, v_{j-1}\rangle.
  \]
Adding these two equalities we obtain
 \begin{equation}
 \label{induction}
  (\lambda+\overline\sigma)\langle Ku_i,v_j\rangle= - \langle Ku_{i-1},v_j\rangle
  - \langle Ku_i, v_{j-1}\rangle.
 \end{equation}
The first step of the induction argument is trivial,
  \[
  \langle Ku_0,v_j\rangle = \langle Ku_i,v_0\rangle = 0,\quad \forall \
  i=1,\dots,n,\ j=1,\dots,m,
  \]
since $u_0=v_0=0$,
and we then conclude using the equality (\ref{induction}) and the hypothesis $\lambda+\overline\sigma\not=0$.
\end{Proof}

The following result which holds for the unstable eigenvalues of $JL$ is an immediate consequence of Lemma \ref{l:1}.

\begin{Corollary}\label{c:1}
  Under the assumptions of Hypothesis~\ref{h:1}, if $u$ belongs to the spectral subspace $E_u$ associated to the unstable spectrum $\sigma_u(JL)$ of $JL$, then
  \[
\langle Ku,u\rangle = 0.
\]
\end{Corollary}

We can now state the abstract counting result as follows.

\begin{Theorem}\label{t:1}
Under the assumptions in Hypothesis~\ref{h:1} the following properties hold.
\begin{enumerate}
\item The number $\mathrm n_u(JL)$ of unstable eigenvalues of the operator $JL$ (counted with algebraic multiplicities) and the number $\mathrm n_{sc}(K)= \mathrm n_s(K)+\mathrm n_c(K)$ of nonpositive eigenvalues of the self-adjoint operator $K$ (counted with multiplicities) satisfy
  \begin{equation*}
\mathrm n_u(JL)\leq \mathrm n_{sc}(K).
\end{equation*}
\item If, in addition, the kernel of the operator $K$ is contained in the kernel of the operator $JL$, then
  \begin{equation}
  \label{e:count-kernel}
  \mathrm n_u(JL)\leq \mathrm n_{s}(K).
  \end{equation}
\end{enumerate}
\end{Theorem}

\begin{Proof}
{\it (i)}  According to Hypothesis~\ref{h:1}~(iii), the spectral subset $\sigma_{sc}=\sigma_s(K)\cup\sigma_c(K)$ of the self-adjoint operator $K$ is a finite set, and  we can consider the corresponding spectral decomposition of the Hilbert space $\mathcal H$,
  \begin{equation}\label{e:fscu}
  \mathcal H = F_{sc} \oplus F_u,\quad
  \sigma(K\big|_{ F_{sc}})= \sigma_{sc}(K),\quad
   \sigma(K\big|_{ F_{u}})= \sigma_{u}(K).
 \end{equation}
We denote by $P_{sc}$ the unique spectral projection onto $F_{sc}$. In particular, \[\mathrm{dim}(F_{sc})= \mathrm n_s(K)+\mathrm n_c(K)=n_{sc}(K),\] and
 \begin{equation}\label{e:kuu>}
\langle Ku,u\rangle>0,\quad\forall\  u\in F_u\setminus\{0\}.
 \end{equation}
Similarly, according to  Hypothesis~\ref{h:1}~(iv), we consider the spectral subspace $E_u$  associated to the unstable spectrum $\sigma_u(JL)$ of $JL$, for which we have that $\mathrm{dim}(E_u)=n_{u}(JL)$.

We claim that the restriction to $E_u$ of the spectral projection $P_{sc}$ is an injective operator $P_{sc}\big|_{E_u}:E_u\to F_{sc}$. Indeed, assume that $P_{sc}u=0$, for some $u\in E_u$. Then $u\in F_u$ and $\langle Ku,u\rangle>0$,
if $u\not=0$, by \eqref{e:kuu>}. On the other hand, according to Corollary~\ref{c:1},  $\langle Ku,u\rangle=0$, since $u\in E_u$. Consequently, $u=0$ which proves the claim. Since $F_{sc}$ is a finite-dimensional space, by Hypothesis~\ref{h:1}~(iii), the injectivity of $P_{sc}\big|_{E_u}$ implies that
\[
\mathrm{dim}(E_u)=n_{u}(JL) \leq \mathrm{dim}(F_{sc})=n_{sc}(K),
\]
and proves the first part of the theorem.

{\it (ii)} In the arguments above, we now replace the spectral decomposition \eqref{e:fscu} of $\mathcal H$ by
 \[
  \mathcal H = F_{s} \oplus F_{cu},\quad
  \sigma(K\big|_{ F_{s}})= \sigma_{s}(K),\quad
   \sigma(K\big|_{ F_{cu}})= \sigma_{cu}(K),
 \]
 and work with the spectral projection $P_{s}$ onto $F_{s}$, instead of $P_{sc}$. In this case, the restriction $P_{s}\big|_{E_u}:E_u\to F_{s}$ is injective. Indeed, assume that $P_{s}u=0$, for some $u\in E_u$. Then $u\in F_{cu}$ and by Corollary~\ref{c:1} we have that $\langle Ku, u\rangle=0$. Together with the inequality \eqref{e:kuu>} this implies that $u$ belongs to the kernel $F_c$ of $K$, and hence to the kernel of $JL$, by hypothesis. We conclude that $u=0$, which proves the injectivity of $P_{s}\big|_{E_u}$. This latter property implies the inequality \eqref{e:count-kernel} and completes the proof of the theorem.
\end{Proof}

The following corollary is a particular case of Theorem~\ref{t:1} for nonnegative operators $K$.

\begin{Corollary}
\label{cor-stability}
Under the assumptions of Hypothesis~\ref{h:1}, further assume that $K$ is a nonnegative operator. Then  $\mathrm n_u(JL)\leq \mathrm n_c(K)$. If in addition the kernel of $K$ is contained in the kernel of $JL$, then $\mathrm n_u(JL)=0$, and the spectrum of $JL$ is purely imaginary.
\end{Corollary}

\begin{Remark}
The particular case of Theorem~\ref{t:1} with $K=L$ recovers the classical counting result showing that $n_{u}(JL) \leq n_{s}(L)$. More refined versions of this result are available in the literature in which, under different additional assumptions, the inequality is replaced by an 
equality (e.g., see \cite{CP, HK, KKS}).
The difference $n_{s}(L)-n_{u}(JL)$ is shown to be given by the number of purely imaginary eigenvalues
of $JL$ which have a negative Krein signature. We expect that such results can be  extended to the present setting by introducing for the purely imaginary eigenvalues of $JL$ a Krein signature relative to the operator $K$.
\end{Remark}

\section{Transverse stability of periodic waves in the KP-II equation}

As an application of the general result in Theorem \ref{t:1}, we discuss the transverse stability of periodic traveling waves in the KP-II equation \eqref{KP}. Here we formulate the transverse spectral and linear stability problems and state the main results. We prove these results in Section~4.

\subsection{Transverse spectral stability}

One-dimensional periodic traveling waves of the KP-II equation (\ref{KP}) are
solutions of the KdV equation (\ref{kdv}) of the form $u(x,t) = \phi_c(x+ct)$, with $\phi_c$ a periodic function and $c$ a constant speed of propagation.
In Section~4, we recall some well-known properties of these periodic traveling waves which are needed for our analysis.
Without loss of generality, we can restrict to $2\pi$-periodic and even solutions $\phi_c$, for $c>1$, as given by Proposition~\ref{prop-wave-explicit}.

In a coordinate system moving with the speed $c$ of the periodic traveling wave,
the corresponding linearization of the KP-II equation~\eqref{KP} is given by
\begin{equation}
\label{KP-linear}
(w_t + w_{xxx} + cw_x + 6 (\phi_c(x)w)_x )_x + w_{yy} = 0,
\end{equation}
in which, for notational simplicity, we denoted by $x$ the variable $x+ct$ from the KP-II equation \eqref{KP}.
Following the transverse spectral stability approach in \cite{Haragus}, we consider solutions of the form
\[
w(x,y,t)=e^{\lambda t+ip y}W(x),
\]
with $W$ satisfying the differential equation
\[
\lambda W_{x}+W_{xxxx}+cW_{xx} + 6(\phi_c(x)W)_{xx} - p^2 W=0.
\]
The left hand side of this equation defines a linear differential operator with $2\pi$-periodic coefficients
\[
\mathcal A_{c,p}(\lambda) = \lambda\partial_x+\partial_{x}^4+
c\partial_{x}^2+6\partial_{x}^2(\phi_c(x)\,\cdot)-p^2,
\]
and the spectral stability problem is concerned with the invertibility
of this operator, for certain values of $p$ and in a suitable function space. The periodic wave
$\phi_c$ is spectrally stable if $\mathcal A_{c,p}(\lambda)$ is invertible for any
$\lambda\in\C$ with $\Re\lambda>0$,
and unstable otherwise. The type of the perturbations determines the choice of the underlying function space and the values of $p$. Here we consider general bounded two-dimensional
perturbations of the periodic wave, and we therefore assume that $\mathcal A_{c,p}(\lambda)$ acts in $C_b(\R)$, the Banach space of uniformly bounded continuous
functions on $\R$, and consider any real number~$p$.

The particular case $p=0$ corresponds to one-dimensional perturbations of the periodic wave which do not depend upon the transverse variable $y$. The dynamics of such perturbations is better described by the KdV equation rather than the KP equation. In particular, the operator $\mathcal A_{c,0}(\lambda)$ obtained using the KP equation has an unnecessary factor $\partial_x$, which is also a noninvertible operator. It is therefore more appropriate to replace in this case the operator  $\mathcal A_{c,0}(\lambda)$ by the one given by the KdV equation,
\[
\widetilde{\mathcal A}_{c,0}(\lambda) = \lambda+\partial_{x}^3+
c\partial_{x}+6\partial_{x}(\phi_c(x)\,\cdot),
\]
for which the invertibility question is equivalent to the one of studying the spectrum of the operator
\[
\widetilde{\mathcal B}_{c,0} = -\partial_{x}^3-
c\partial_{x}-6\partial_{x}(\phi_c(x)\,\cdot).
\]
The results in \cite{b2} (see also \cite{HK} for the case of small-amplitude waves) imply that the spectrum of this operator is purely imaginary, hence showing spectral stability with respect to one-dimensional perturbations.

Truly two-dimensional perturbations correspond to $p\not=0$.
Since spectra of differential operators with periodic coefficients acting in $C_b(\R)$ are typically continuous,
the Hypothesis~\ref{h:1}(iv), which requires point spectra, is not satisfied with this choice of the function space.
In order to overcome this difficulty, we use first a Bloch decomposition, based on Floquet theory,
showing that the operator $\mathcal A_{c,p}(\lambda)$ is invertible in  $C_b(\R)$ if and only if the operators
\begin{equation*}
\mathcal A_{c,p}(\lambda,\gamma) = \lambda(\partial_x+i\gamma)+(\partial_{x}+i\gamma)^4+
c(\partial_{x}+i\gamma)^2+6(\partial_{x}+i\gamma)^2(\phi_c(x)\,\cdot)-p^2,
\end{equation*}
are invertible in the space $L^2_{per}(0,2\pi)$ of square-integrable $2\pi$-periodic functions,
for any $\gamma\in[0,1)$ (e.g., see \cite{Haragus}).
At this point, we distinguish the cases $\gamma\not=0$ and $\gamma=0$.

For any $\gamma\in(0,1)$, the operator $\partial_x+i\gamma$ has a bounded inverse in
$L^2_{per}(0,2\pi)$, so that $\mathcal A_{c,p}(\lambda,\gamma)$ is invertible if and only if $\lambda$ belongs to the resolvent set of the operator
\begin{equation}\label{e:bcp}
\mathcal B_{c,p}(\gamma) = -(\partial_{x}+i\gamma)^3
-c(\partial_{x}+i\gamma)-6(\partial_{x}+i\gamma)(\phi_c(x)\,\cdot)+p^2(\partial_{x}+i\gamma)^{-1},
\end{equation}
which is a closed operator in  $L^2_{per}(0,2\pi)$ with domain  $H^3_{per}(0,2\pi)$.
Consequently, our problem is reduced to that of studying the spectrum of $\mathcal B_{c,p}(\gamma)$,
which is an operator with compact resolvent, hence with point spectrum consisting of isolated eigenvalues
with finite algebraic multiplicities, only. Moreover, $\mathcal B_{c,p}(\gamma)$ has the $JL$ product structure in the previous section,
\begin{equation}\label{e:bcpgamma}
\mathcal B_{c,p}(\gamma) =J(\gamma) L_{c,p}(\gamma),
\end{equation}
with
\begin{equation}\label{e:JLcp}
  J(\gamma) =  (\partial_{x}+i\gamma),\quad
L_{c,p}(\gamma) = - (\partial_{x}+i\gamma)^2
-c-6\phi_c(x) + p^2(\partial_{x}+i\gamma)^{-2}.
\end{equation}
It is not difficult to check that the operators $J(\gamma)$ and $L_{c,p}(\gamma)$ satisfy
the properties required by the Hypothesis~\ref{h:1}.

In contrast, for $\gamma=0$, the operator $\partial_x$ is not invertible in  $L^2_{per}(0,2\pi)$.
However, for $p\not=0$, any function in the kernel of  $\mathcal A_{c,p}(\lambda,0)$ has zero mean,
so that the invertibility of  $\mathcal A_{c,p}(\lambda,0)$ in $L^2_{per}(0,2\pi)$ is equivalent to the invertibility
of $\mathcal A_{c,p}(\lambda,0)$ in the invariant subspace $\dot{L}^2_{per}(0,2\pi)$ of functions with zero mean.
In this subspace, $\partial_x$ has a bounded inverse, and  $\mathcal A_{c,p}(\lambda,0)$ is invertible
if and only if $\lambda$ belongs to the resolvent set of the operator $\mathcal B_{c,p}(0)$ defined
in \eqref{e:bcp} for $\gamma=0$. We point out that  $\dot{L}^2_{per}(0,2\pi)$ is an invariant subspace
for the operators $\mathcal B_{c,p}(0)$ and $J(0)$ but not for $L_{c,p}(0)$. Therefore, in the subsequent analysis,
we replace, when needed, the operator  $L_{c,p}(0)$ by the projected operator $\Pi_0L_{c,p}(0)$,
where $\Pi_0:{L}^2_{per}(0,2\pi)\to\dot{L}^2_{per}(0,2\pi)$ is the standard orthogonal projection
on the nonzero Fourier modes. For notational simplicity, we denote the projected operator $\Pi_0 L_{c,p}(0)$
also by $L_{c,p}(0)$, and refer to it as the restriction of  $L_{c,p}(0)$  to $\dot{L}^2_{per}(0,2\pi)$.

Summarizing, we can restrict our analysis to the case of truly two-dimensional bounded perturbations, $p\not=0$.
Nevertheless, the existing results for the limit case $p=0$ from \cite{b1} will play a key role in the subsequent proofs.
The arguments above show that the question of transverse spectral stability for a periodic wave $\phi_c$ reduces to the study of the (point) spectrum of the operators $\mathcal B_{c,p}(\gamma)$. For this spectral analysis, we apply the general counting result in Corollary~\ref{cor-stability} with $J=J(\gamma)$, $L=L_{c,p}(\gamma)$, and  suitably chosen operators $K=K_{c,p}(\gamma)$, which are nonnegative. These operators are constructed in Sections 4.2 and 4.3 below.
We obtain the following theorem showing transverse spectral stability.

\begin{Theorem}\label{t:spectral}
Consider a periodic traveling wave  $\phi_c$ of the KdV equation \eqref{kdv}
with the properties given in Proposition~\ref{prop-wave-explicit} below. For every $p\not=0$, the following properties hold.
\begin{enumerate}
\item The linear operator $\mathcal B_{c,p}(\gamma) =J(\gamma) L_{c,p}(\gamma)$ defined in \eqref{e:bcpgamma}--\eqref{e:JLcp}, acting in  $L^2_{per}(0,2\pi)$, when $\gamma\in(0,1)$, and in  $\dot{L}^2_{per}(0,2\pi)$, when $\gamma=0$,  has purely imaginary spectrum, for any $\gamma\in[0,1)$.
\item The linear operator $\mathcal A_{c,p}(\lambda)$ is invertible in $C_b(\R)$, for any $\lambda\in\C$ with $\Re\lambda>0$.
\end{enumerate}
Consequently, the periodic traveling wave  $\phi_c$ is transversely spectrally stable with respect to two-dimensional bounded perturbations.
\end{Theorem}

We prove the first part of this theorem in Section~4.4. The second part is an immediate consequence of the arguments above.

\begin{Remark}\label{r:L}
As explained in \cite{Haragus}, in the case of small-amplitude limit, $c \to 1$,
the spectral properties of $L_{c,p}(\gamma)$ for $p \neq 0$  are not good enough to conclude
on spectral stability using the classical counting criterion (with $K=L$). Indeed, with Fourier series, we find that the spectrum of the limit operator $L_{1,p}(\gamma)$ from (\ref{e:JLcp})  is given by
\begin{eqnarray*}
\sigma({L}_{1,p}(\gamma)) = \left\{ k^2 - 1 -{p^2}{k^{-2}}\;;\; k=\gamma+n,\ n\in\Z,\ \gamma+n\not=0\right\}.
\end{eqnarray*}
Since the map $k\mapsto k^2-1-p^2k^{-2}$ is negative for $k^2\leq(1+\sqrt{1+4p^2})/2$, the operator has an increasing number of negative eigenvalues as $p\to\infty$. This property remains true for values of $c$ close to $1$, hence making difficult to conclude on the absence of unstable eigenvalues for the operator $J(\gamma) L_{c,p}(\gamma)$ for any~$p$.
\end{Remark}

\subsection{Transverse linear stability}

The positivity properties of the operators  $K=K_{c,p}(\gamma)$ used in our spectral stability analysis, also allows us to prove a transverse linear stability result. However, this latter result is restricted to doubly periodic perturbations, which are subharmonic with zero mean in the direction of propagation $x$ and have an arbitrary, but fixed, period in the transverse direction $y$.

Restricting to periodic perturbations which have zero mean in $x$, we rewrite the linearized equation \eqref{KP-linear} as an evolutionary problem
\begin{equation}\label{e:linKP}
w_t = \mathcal B_cw,
\end{equation}
in which $\mathcal B_c$ is a differential operator with $2\pi$-periodic coefficients having a $JL$-product structure, more precisely,
\begin{equation}
  \mathcal B_c = JL_c, \quad
\label{JLc}
J=\partial_x,\quad L_c =- \partial_{x}^2 - c - 6 \phi_c(x) - \partial_x^{-2}\partial_{y}^2.
\end{equation}
Here, the operator $\mathcal B_c$ is well-defined and closable in the space of
locally square-integrable functions on $\R^2$ which are $2\pi N$-periodic and
have zero mean in $x$, for some $N\in\N$, and are $2\pi/p$-periodic in $y$,
for some fixed wave number $p$. We denote this space by $\dot L^2(N,p)$. In this space, the operators $J$ and $L_c$ are skew- and self-adjoint operators, respectively.

The key observation in our linear stability proof is that the existence of
a self-adjoint operator $K_c$ satisfying the commutativity property
\begin{equation}\label{e:lckc}
L_cJK_c=K_cJL_c,
\end{equation}
just as the ones in Hypothesis~\ref{h:1}(ii),
implies that the associated quadratic form $\langle K_c\cdot,\cdot\rangle$
is constant along suitable solutions to the linearized equation \eqref{e:linKP},
hence it acts as a Lyapunov functional. Indeed, a simple formal calculation gives
\begin{eqnarray*}
\frac{d}{dt}\langle K_cw,w\rangle  =  \langle K_c JL_c w, w \rangle + \langle K_c w,JL_cw\rangle
 =  \langle K_c JL_c w, w \rangle - \langle L_cJK_cw,w\rangle = 0.
\end{eqnarray*}
This calculation becomes rigorous for appropriately regular solutions.
Thus, for suitable solutions $w(t)$ to the linearized equation \eqref{e:linKP}, we have
\begin{equation}\label{e:kwt}
\langle K_cw(t),w(t)\rangle =
\langle K_cw(0),w(0)\rangle,\quad \forall\ t\in\R.
\end{equation}
If the operator $K_c$ is coercive in some norm, then the solutions $w(t)$ to the linearized equation \eqref{e:linKP}
stay bounded in this norm for all times, which then implies linear stability.

The transverse linear stability result is obtained in the energy space for the quadratic form (\ref{e:kwt}),
which coincides with the Hilbert space $H^{2,1}(N,p)$ defined by
\begin{equation*}
H^{2,1}(N,p) = \{w\in \dot L^2(N,p) : \quad  w_x,w_{xx},w_y\in \dot L^2(N,p)\},
\end{equation*}
and equipped with the standard norm denoted by $\|\cdot\|_{2,1}$. The following theorem is
proved in Section~4.5.

\begin{Theorem}\label{t:linear}
Consider a periodic traveling wave $\phi_c$ of the KdV equation \eqref{kdv}
with the properties given in Proposition~\ref{prop-wave-explicit} below. For any $N\in\N$ and any positive $p\in\R$,
there exists a constant $C_{N,p}$
 such that any solution $w\in C^1(\R,H^{2,1}(N,p))$ of the linearized equation \eqref{e:linKP} satisfies the inequality
\begin{equation}
\label{bound-theorem}
\|w(t) - a(t) \partial_x \phi_c \|_{2,1}\leq C_{N,p}\|w(0)\|_{2,1},
\quad |a'(t)| \leq C_{N,p},
\end{equation}
where $a(t)$ represents the orthogonal projection of the solution on the derivative $ \partial_x \phi_c$ of the periodic wave,
\[
a(t) = \langle w(t), \partial_x \phi_c \rangle.
\]
Consequently, the periodic traveling wave is transversely linearly stable with respect to doubly periodic perturbations in $H^{2,1}(N,p)$.
\end{Theorem}

\begin{Remark}
  \begin{enumerate}
  \item Due to the translation invariance of the KP-II equation, the derivative  $ \partial_x \phi_c$ of the periodic wave belongs to the kernel of the linearized operator $\mathcal B_c$. As we shall see later, it also belongs to the kernel of the operator $K_c$, which is only coercive on the subspace orthogonal to   $ \partial_x \phi_c$. This explains the presence of the term $a(t) \partial_x \phi_c$ in the first estimate in \eqref{bound-theorem}. Furthermore, the linearized operator  $\mathcal B_c$ has a generalized kernel with one, at least, $2\times2$ Jordan block. This  explains a possible linear growth of $a(t)$, as indicated by the second inequality in \eqref{bound-theorem}. The estimates in \eqref{bound-theorem} are the linear counterpart of a standard nonlinear orbital stability result claiming that, as expected in the presence of translational invariance, solutions stay close to the orbit  $\{ \phi_c(\cdot + x_0) \}_{x_0 \in \mathbb{R}}$ of the periodic traveling wave.
  \item We do not discuss here the initial value problem for the linearized equation \eqref{e:linKP}, and hence the question of existence of solutions $w\in C^1(\R,H^{2,1}(N,p))$. However, on the basis of semigroup theory, one expects that for initial data $w(0)\in H^{5,3}(N,p)$ a unique solution of \eqref{e:linKP} exists which satisfies $w\in C^1(\R,H^{2,1}(N,p))\cap  C^0(\R,H^{5,3}(N,p))$, where the space $H^{5,3}(N,p)$ is defined similarly to $H^{2,1}(N,p)$.
  \end{enumerate}
\end{Remark}

\section{Proofs of Theorems \ref{t:spectral} and \ref{t:linear}}

Here we prove the stability results in Theorems \ref{t:spectral} and \ref{t:linear}.
We recall some well-known properties of the periodic traveling waves of the KdV equation (\ref{kdv})
in Section~4.1. In Sections~4.2 and~4.3, we construct the operators $K=K_{c,p}(\gamma)$ and discuss
their positivity properties. We conclude with the proofs of the two theorems in Sections~4.4 and~4.5.

\subsection{One-dimensional periodic traveling waves}

Periodic traveling waves of the KdV equation (\ref{kdv}) are
solutions of the form $u(x,t) = v(x+ct)$, with $v$ a periodic function in its argument. Due to the Galilean invariance, one can integrate the resulting third-order
differential equation for $v$ with zero integration constant and obtain $v$
from the second-order differential equation
\begin{equation}
\label{ode-kdv}
v''(x) + c v(x) + 3 v^2(x) = 0.
\end{equation}
Without loss of generality, due to scaling and translation invariances, we scale the period of the periodic traveling wave to $2\pi$,
translate the wave profile $v$ to be even in $x$, and so restrict to $2\pi$-periodic even
solutions to the differential equation (\ref{ode-kdv}).
A complete characterization of these periodic waves is available in terms of Jacobi elliptic
functions (e.g., see \cite{b1}). The following proposition specifies this explicit result.

\begin{Proposition}
\label{prop-wave-explicit}
For every $c > 1$, the
differential equation (\ref{ode-kdv}) possesses a unique $2\pi$-periodic even solution $\phi_c$ which satisfies $\phi_c(0) > 0$ and is given by
\begin{equation}
\label{wave-explicit}
\phi_c(x) = \frac{2 K^2(k)}{3 \pi^2} \left[  1 - 2k^2 - \sqrt{1 - k^2 + k^4} + 3 k^2 {\rm cn}^2\left(\frac{K(k)}{\pi} x; k\right) \right].
\end{equation}
Here ${\rm cn}$ is the Jacobi elliptic function,
$K(k)$ is a complete elliptic integral, and the elliptic modulus $k \in (0,1)$ parameterizes
the speed parameter $c$ by
\begin{equation}
\label{speed-explicit}
c = \frac{4 K^2(k)}{\pi^2} \sqrt{1 - k^2 + k^4}.
\end{equation}
\end{Proposition}

\begin{Proof}
It follows from the explicit expressions involving Jacobi elliptic functions (e.g., see \cite{b1}),
that the function
\begin{equation*}
u(\xi) = 2 k^2 {\rm cn}^2(\xi;k), \quad k \in (0,1),
\end{equation*}
is a $2 K(k)$-periodic solution of the second-order differential equation
\begin{equation}\label{second-order-u}
u''(\xi) + 4 (1-2k^2) u(\xi) + 3 u^2(\xi) = 4 k^2 ( 1 - k^2).
\end{equation}
In order to remove the constant term from the right-hand side of equation (\ref{second-order-u}),
and normalize the period of $u$ to $2 \pi$, we use the scaling and shift transformation
\begin{equation}
\label{wave-transformation}
\phi_c(x) = \frac{K^2(k)}{\pi^2} \left[ A(k) + u\left(\frac{K(k)}{\pi} x\right) \right],
\end{equation}
and take
\begin{equation}
\label{speed-transformation}
c = \frac{K^2(k)}{\pi^2} \left[ 4 (1 - 2k^2) - 6 A(k) \right],
\end{equation}
where $A(k)$ is a solution of quadratic equation
\begin{equation}
\label{wave-quadratic}
3 A^2 - 4(1-2k^2) A - 4 k^2 (1-k^2) = 0,
\end{equation}
satisfying $A(0) = 0$. Solving the quadratic equation (\ref{wave-quadratic}), we
obtain
\begin{equation}
\label{A-def}
A(k) = \frac{2}{3} \left[ 1 - 2k^2 - \sqrt{1 - k^2 + k^4} \right].
\end{equation}
Substituting (\ref{A-def}) into (\ref{wave-transformation}) and (\ref{speed-transformation}),
we obtain (\ref{wave-explicit}) and (\ref{speed-explicit}).
\end{Proof}

\begin{Remark}
\label{remark-Stokes}
As $k \to 0$, the explicit solution given by (\ref{wave-explicit}) and (\ref{speed-explicit})
recovers the Stokes expansion for small-amplitude periodic waves,
\begin{equation}
\label{stokes-wave}
\phi_c(x) = a \cos(x) + \frac{a^2}{2} \left[ \cos(2x) - 3 \right] + \mathcal{O}(a^3), \quad
c = 1 + \frac{15}{2} a^2 + \mathcal{O}(a^4),
\end{equation}
where $a = k^2/4 + \mathcal{O}(k^4)$ is the projection to the first Fourier mode.
Note that $c > 1$ follows from (\ref{speed-explicit}) for every $k \in (0,1)$.
\end{Remark}

\subsection{Construction of  commuting operators $\boldsymbol{M_{c,p}(\gamma)}$}

We start by constructing a self-adjoint operator $M_{c,p}(\gamma)$ which satisfies the commutativity condition
\eqref{e:jljk} in Hypothesis~\ref{h:1}(ii). For notational simplicity, we restrict in the following arguments to
the case $\gamma=0$ and take
\begin{equation}
\label{operator-L}
J =  \partial_{x},\quad
L_{c,p} = - \partial_{x}^2
-c-6\phi_c(x) + p^2\partial_{x}^{-2}.
\end{equation}
For $\gamma\not=0$, the operators $M_{c,p}(\gamma)$ are easily obtained from the resulting operator $M_{c,p}$ by formally replacing the derivative $\partial_x$ with $\partial_x+i\gamma$.

We search for a self-adjoint operator $M_{c,p}$
which satisfies the commutativity condition (\ref{e:jljk}) in Hypothesis~\ref{h:1}~(ii).
As in Remark~\ref{r:1}~(i), we write the commutativity condition in the form
\begin{equation}
\label{commutability}
L_{c,p} \partial_x M_{c,p} = M_{c,p} \partial_x L_{c,p}.
\end{equation}
For the purpose of symbolic computations, we write
\begin{equation}
\label{L-expansion}
L_{c,p} = L_{\rm KdV} + p^2 L_{\rm KP},
\end{equation}
where
\begin{equation}
\label{L-explicit}
L_{\rm KdV} = - \partial_x^2 - c - 6 \phi_c(x), \quad L_{\rm KP} = \partial_x^{-2},
\end{equation}
and similarly,
\begin{equation}
\label{M-expansion}
M_{c,p} = M_{\rm KdV} + p^2 M_{\rm KP},
\end{equation}
where $M_{\rm KdV}$ and $M_{\rm KP}$ are the operators to be found.

The case $p = 0$ corresponds to the KdV equation for which the operator $M_{\rm KdV}$
has been constructed in \cite{b1}. We briefly recall this construction here.
The operators $L_{\rm KdV}$ and $M_{\rm KdV}$
are related to linearized equations of the KdV hierarchy.
Formally, the second-order differential equation (\ref{ode-kdv}) is the Euler--Lagrange
equation for the energy functional
\begin{equation}
\label{energy-S}
S_c(u) = E(u) - c Q(u),
\end{equation}
where $E(u)$ and $Q(u)$ are the Hamiltonian and momentum, respectively, of the KdV equation (\ref{kdv}) given by
\begin{equation*}
E(u) = \int \left[ u_x^2 - 2 u^3\right] dx,
\quad
Q(u) = \int u^2 dx .
\end{equation*}
The higher-order energy functional
of the KdV equation takes the form
\begin{equation}
\label{higher-order-energy-KdV}
H(u) = \int \left[ u_{xx}^2 - 10 u u_x^2 + 5 u^4 \right] dx.
\end{equation}
To obtain $M_{\rm KdV}$, we observe that a solution
of the second-order differential equation (\ref{ode-kdv}) is also
a critical point of the higher-order energy functional
\begin{equation}
\label{energy-R}
R_c(u) = H(u) - c^2 Q(u) + 2 IC(u),
\end{equation}
where $C(u) = \int u\, dx$ is the Casimir-type functional, which does not contribute to the second variation,
whereas $I$ is the first-order invariant for the second-order differential equation (\ref{ode-kdv}) given by
\begin{equation*}
I = \left( \frac{d v}{d x} \right)^2 + c v^2 + v^3 = {\rm const}.
\end{equation*}
By computing the Hessian operator of $R_c(u)$
at the periodic wave $\phi_c$, we obtain the linear operator
\begin{equation}
\label{M-explicit}
M_{\rm KdV} = \partial_x^4 + 10 \partial_x \phi_c(x) \partial_x - 10 c \phi_c(x) - c^2,
\end{equation}
and straightforward symbolic computations confirm that
\begin{equation*}
L_{\rm KdV} \partial_x M_{\rm KdV} - M_{\rm KdV} \partial_x L_{\rm KdV} = 0.
\end{equation*}

Next, we are looking for $M_{\rm KP}$ from the commutativity condition
\begin{equation}
\label{commutability-p-2}
L_{\rm KdV} \partial_x M_{\rm KP} - M_{\rm KP} \partial_x L_{\rm KdV} =
M_{\rm KdV} \partial_x L_{\rm KP} - L_{\rm KP} \partial_x M_{\rm KdV},
\end{equation}
which corresponds to the order $\mathcal{O}(p^2)$ obtained from (\ref{commutability}), (\ref{L-expansion}),
and (\ref{M-expansion}). From the explicit expressions (\ref{L-explicit}) and (\ref{M-explicit}), we find
the right-hand side of (\ref{commutability-p-2}),
\begin{equation*}
M_{\rm KdV} \partial_x L_{\rm KP} - L_{\rm KP} \partial_x M_{\rm KdV} =
10 v'(x) + 10 c \left( \partial_x^{-1} \phi_c(x) - \phi_c(x) \partial_x^{-1} \right).
\end{equation*}
On the other hand, the left-hand side of (\ref{commutability-p-2}) is given by the operator
\begin{eqnarray*}
\nonumber
& \phantom{t} & L_{\rm KdV} \partial_x M_{\rm KP} - M_{\rm KP} \partial_x L_{\rm KdV} \\
& = & M_{\rm KP} \partial_x^3 - \partial_x^3 M_{\rm KP}
+ c \left( M_{\rm KP} \partial_x - \partial_x M_{\rm KP}\right)
+ \, 6 \left(M_{\rm KP} \partial_x \phi_c(x) - \phi_c(x) \partial_x M_{\rm KP} \right). 
\end{eqnarray*}
By using symbolic computations, again, we obtain that the operator
\begin{equation}
\label{M-explicit-p-2}
M_{\rm KP} = \frac{5}{3} \left(1 + c \partial_x^{-2} \right)
\end{equation}
is a solution of the linear equation (\ref{commutability-p-2}). Moreover,
since $L_{\rm KP}$ and $M_{\rm KP}$ in (\ref{L-explicit}) and (\ref{M-explicit-p-2}) are operators with constant coefficients,
the commutativity condition (\ref{commutability}) at order $\mathcal{O}(p^4)$ is satisfied identically:
\begin{equation*}
L_{\rm KP} \partial_x M_{\rm KP} - M_{\rm KP} \partial_x L_{\rm KP} = 0.
\end{equation*}
Thus, the commutativity condition (\ref{commutability}) is satisfied at all orders with the operator
$M_{c,p}$ given  by (\ref{M-expansion}), (\ref{M-explicit}), and (\ref{M-explicit-p-2}),
or explicitly, by
\begin{equation}
\label{operator-M-equivalent}
M_{c,p} = \partial_x^4 + 10 \partial_x \phi_c(x) \partial_x - 10 c \phi_c(x) - c^2 + \frac{5}{3} p^2 \left(1 + c \partial_x^{-2} \right).
\end{equation}

Finally, by replacing $\partial_x$ with $\partial_x+i\gamma$ in \eqref{operator-M-equivalent} we find
\begin{equation}
\label{e:Mcp}
M_{c,p}(\gamma) = (\partial_x+i\gamma)^4 + 10 (\partial_x+i\gamma) \phi_c(x) (\partial_x+i\gamma) - 10 c \phi_c(x) - c^2 + \frac{5}{3} p^2 \left(1 + c (\partial_x+i\gamma)^{-2} \right).
\end{equation}
This operator is well-defined and self-adjoint in  $L^2_{per}(0,2\pi)$, for any $\gamma\in(0,1)$. For $\gamma=0$, we use the restriction of $M_{c,p}(0)$  to  $\dot L^2_{per}(0,2\pi)$, as explained in Section~3.1 for the operator $L_{c,p}(0)$.

\begin{Remark}
\label{rem-M-property}
For $c=1$, when $\phi_c=0$ and the operators have constant coefficients, we can explicitly compute the spectrum of  $M_{1,p}(\gamma)$ in \eqref{e:Mcp}. We obtain
\begin{eqnarray*}
\sigma({M}_{1,p}(\gamma)) = \left\{  k^4 - 1 + \frac{5 p^2}{3} - \frac{5 p^2}{3 k^2}\;;\; k= \gamma+n,\ n\in\Z,\ \gamma+n\not=0 \right\},
\end{eqnarray*}
from which we conclude that the operators ${M}_{1,p}(\gamma)$ have at least some negative eigenvalues,
just as  ${L}_{1,p}(\gamma)$.
However,  the linear combination $ {M}_{1,p}(\gamma) - 2{L}_{1,p}(\gamma)$ of these two operators
has a nonnegative spectrum,
\begin{eqnarray}
\label{operator-M-positivity}
\sigma({M}_{1,p}(\gamma) - 2{L}_{1,p}(\gamma)) =
\left\{  (k^2-1)^2 + \frac{5 p^2}{3} + \frac{p^2}{3 k^2}\;;\; k= \gamma+n,\ n\in\Z,\ \gamma+n\not=0 \right\}.
\end{eqnarray}
In the next section we show that, by choosing an appropriate linear combination of the operators $M_{c,p}(\gamma)$ and  $L_{c,p}(\gamma)$, this positivity property can be extended to all $c>1$.
\end{Remark}

\subsection{Construction of positive operators $\boldsymbol{K_{c,p,b}(\gamma)}$}

Our construction of a positive linear combination of the operators $M_{c,p}(\gamma)$ and  $L_{c,p}(\gamma)$, relies upon the following result obtained for the KdV equation in \cite{b1}, which corresponds to $p=0$ in our case.

\begin{Proposition}
  \label{prop-positivity}
Consider a periodic traveling wave $\phi_c$ of the KdV equation \eqref{kdv}
with the properties given in Proposition~\ref{prop-wave-explicit}, and  a linear combination of the operators $L_{c,0}$ and $M_{c,0}$ in (\ref{operator-L}) and (\ref{operator-M-equivalent}),
\begin{equation}\label{e:Kcp}
K_{c,0,b} = M_{c,0} - b L_{c,0},
\end{equation}
for some real number $b$. Assume that $L_{c,0}$,  $M_{c,0}$, and $K_{c,0,b}$ act in $L_{per}^2(0,2\pi N)$, the space of
locally square-integrable functions on $\R$ which are $2\pi N$-periodic.
Then, for any $N\in\N$ and $b \in (b_-(c),b_+(c))$, where
\begin{equation}
\label{b-range-equiv}
b_-(c) = \left( \frac{5}{3} + \frac{1-2k^2}{3 \sqrt{1-k^2+k^4}} \right) c, \quad
b_+(c) = \left( \frac{5}{3} + \frac{1+k^2}{3 \sqrt{1-k^2+k^4}} \right) c,
\end{equation}
with $k \in (0,1)$ being the elliptic modulus in Proposition \ref{prop-wave-explicit}, there exists a positive constant $C_{N,c,b}$ such that
\[
\langle  K_{c,0,b} W, W\rangle\geq C_{N,c,b} \|W\|^2,\quad
\forall \  W\in H^2_{per}(0,2\pi N),\ \langle W,\partial_x\phi_c\rangle=0.
\]
Here $\langle\cdot,\cdot\rangle$ denotes the usual hermitian scalar product in $L^2_{per}(0,2\pi N)$ and $\|\cdot\|$ the corresponding norm.
\end{Proposition}

\begin{Proof}
We transfer the result in \cite{b1} to our variables, just as in the proof of Proposition~\ref{prop-wave-explicit}. According to \cite{b1}, the result in the proposition holds for the operator
\begin{eqnarray*}
\widetilde{K}_{\rm KdV} & = & \partial_{\xi}^4 + 10 u(\xi) \partial_{\xi}^2 + 10 u'(\xi) \partial_{\xi} + 10 u''(\xi) + 30 u(\xi)^2 - 16 + 56 k^2 (1-k^2) \\
& & + c_{21} (-\partial_{\xi}^2 - 6 u(\xi) + 8 k^2 - 4),
\end{eqnarray*}
in which $u$ is the $2 K(k)$-periodic solution of the  second-order differential equation \eqref{second-order-u} in the proof of Proposition~\ref{prop-wave-explicit}, and the constant  $c_{21}$, which plays the role of $b$, satisfies
\begin{equation*}
4 (3k^2-2) < c_{21} < 4 (4k^2-2).
\end{equation*}
Transforming variables
through (\ref{wave-transformation}) and (\ref{speed-transformation}), after some computations, we obtain
\begin{eqnarray*}
&&\widetilde{K}_{\rm KdV} = \frac{\pi^4}{K(k)^4} \left[ \partial_x^4 + 10 \partial_x \phi_c(x) \partial_x - 10 c \phi_c(x) - c^2 \right]
+ \frac{\pi^2}{K(k)^2} \left[ c_{21} + 10 A(k) \right]  \; \left[ -\partial_x^2 - 6 \phi_c(x) - c \right].
\end{eqnarray*}
Comparing the expression of $\widetilde{K}_{\rm KdV}$ with $K_{c,0,b}$ given by
(\ref{operator-L}), (\ref{operator-M-equivalent}), and (\ref{e:Kcp}), we obtain
the correspondence between $b$ and $c_{21}$,
\begin{equation*}
b = -\frac{K(k)^2}{\pi^2} \left[ c_{21} + 10 A(k) \right],
\end{equation*}
and the values of $b$ for which the results in the proposition holds,
\begin{equation*}
\frac{4 K^2(k)}{3 \pi^2} \left( 5 \sqrt{1 - k^2 + k^4} + 1 - 2 k^2 \right) < b < \frac{4 K^2(k)}{3 \pi^2} \left( 5 \sqrt{1 - k^2 + k^4} + 1 + k^2 \right).
\end{equation*}
Finally, using the explicit definition of the speed $c$ in (\ref{speed-explicit}), we obtain the formulas in (\ref{b-range-equiv}).
\end{Proof}

\begin{Remark}
\label{rem-Decon}
The result in this proposition has been proved in \cite{b1} by evaluating the quadratic form associated to $\widetilde{K}_{\rm KdV}$ on a complete set of eigenfunctions of the linearized KdV operator. This set of eigenfunctions is known explicitly, due to the integrability of the KdV equation. Recently, in the context of the cubic NLS equation, such a result has been obtained in \cite{gp} by directly estimating the quadratic form, hence without using the knowledge of an explicit set of eigenfunctions. For the KdV equation there is no such direct proof, so far. However, in the case of small-amplitude solutions 
(see the Stokes expansion (\ref{stokes-wave}) in Remark \ref{remark-Stokes}), such a direct proof can be obtained using perturbation arguments, 
just as recently done in \cite{jp} for the reduced Ostrovsky equations.
\end{Remark}

We consider now a linear combination of the operators $M_{c,p}(\gamma)$ and  $L_{c,p}(\gamma)$,
\begin{equation}\label{e:Kcp-gamma}
K_{c,p,b}(\gamma) = M_{c,p}(\gamma) - b L_{c,p}(\gamma),
\end{equation}
for some real number $b$.
As a consequence of the previous proposition we obtain the following result for $p=0$.

\begin{Corollary}[$\boldsymbol{p=0}$]
  \label{cor-positivity}
Consider a periodic traveling wave $\phi_c$ of the KdV equation \eqref{kdv}
with the properties given in Proposition~\ref{prop-wave-explicit}. Then for every $\gamma\in(0,1)$ and
$b \in (b_-(c),b_+(c))$, where $b_-(c)$ and $b_+(c)$ are given by (\ref{b-range-equiv}),
there exists a positive constant $C_{c,0,b}(\gamma)$ such that the linear operator
$K_{c,0,b}(\gamma)$ satisfies the inequality
\begin{equation*}
\langle K_{c,0,b}(\gamma) W, W\rangle\geq C_{c,0,b}(\gamma) \|W\|^2,\quad
\forall \  W\in H^2_{per}(0,2\pi).
\end{equation*}
For $\gamma=0$, the derivative $\partial_x\phi_c$ of the periodic wave belongs to the kernel of
$K_{c,0,b}(0)$, and the inequality holds for any $W\in H^2_{per}(0,2\pi)$ satisfying $\langle W,\partial_x\phi_c\rangle=0$. Here $\langle\cdot,\cdot\rangle$ denotes the usual hermitian scalar product in
$L^2_{per}(0,2\pi)$ and $\|\cdot\|$ the corresponding norm.
\end{Corollary}

\begin{Proof}
For rational numbers $\gamma = j/N\in[0,1)$, the assertion in this corollary is a consequence of Proposition~\ref{prop-positivity}. Indeed, using Floquet decomposition in $x$, we obtain that the  spectrum of the operator $K_{c,0,b}$ acting in $L_{per}^2(0,2\pi N)$ is given by
\[
\sigma(K_{c,0,b}) = \bigcup_{\gamma\in\mathcal I_N} \sigma\left(K_{c,0,b}(\gamma) \right),\quad \mathcal I_N = \left\{\frac jN\;;\; j=0,\dots,N-1\right\},
\]
where the operators $K_{c,0,b}(\gamma)$ act in $L^2_{per}(0,2\pi)$. Then the result in Proposition~\ref{prop-positivity} implies that  the operators  $K_{c,0,b}(\gamma)$ are positive for any rational number $\gamma = j/N\in(0,1)$, and that for $\gamma=0$ they are nonnegative and have a one-dimensional kernel spanned by $\partial_x\phi_c$. Consequently, the result holds for any rational number $\gamma\in \Q\cap[0,1)$. Finally, the density of $\Q$ in $\R$ together with a standard perturbation argument shows that the result holds for any $\gamma\in[0,1)$, which proves the corollary.
\end{Proof}

We can now state the positivity result for the operators $K_{c,p,b}(\gamma)$ in \eqref{e:Kcp-gamma},
for $p\not=0$. These operators act in $L^2_{per}(0,2\pi)$ when $\gamma\in(0,1)$, and are restricted to
$\dot L^2_{per}(0,2\pi)$ when $\gamma=0$.

\begin{Lemma}[$\boldsymbol{p\not=0}$]
\label{lemma-positivity}
Consider a periodic traveling wave $\phi_c$ of the KdV equation \eqref{kdv}
with the properties given in Proposition~\ref{prop-wave-explicit}. Assume that $p\not=0$.
Then, for any $\gamma\in(0,1)$ and $b\in(b_0(c),b_+(c))$, where  $b_0(c)=\max\{5c/3,b_-(c)\}$ and $b_{\pm}(c)$ are given by (\ref{b-range-equiv}),  there exists a positive constant $C_{c,p,b}(\gamma)$
such that the linear operator $K_{c,p,b}(\gamma)$
defined in \eqref{e:Kcp-gamma}, satisfies the inequality
\[
\langle  K_{c,p,b}(\gamma) W, W\rangle\geq C_{c,p,b}(\gamma) \|W\|^2,\quad
\forall \  W\in H^2_{per}(0,2\pi).
\]
For $\gamma=0$, the same property holds for  $W\in H^2_{per}(0,2\pi)\cap \dot L^2_{per}(0,2\pi)$.
\end{Lemma}

\begin{Proof}
We rewrite
\begin{equation*}
K_{c,p,b}(\gamma) = K_{c,0,b}(\gamma) + \frac{5}{3} p^2 - \left( b-\frac{5c}{3}\right)  p^2 (\partial_x+i\gamma)^{-2}.
\end{equation*}
For any $b>5c/3$, the last two terms in the right hand side of this equality define a positive operator.
Combined with the result in Corollary~\ref{cor-positivity}, this proves the lemma.
\end{Proof}

\subsection{Proof of Theorem~\ref{t:spectral}}

Theorem~\ref{t:spectral}~(i) is a consequence of the general result in
Corollary~\ref{cor-stability}. Indeed,  take $K_{c,p,b}(\gamma)$ with some $b\in(b_0(c),b_+(c))$,
as constructed in Lemma~\ref{lemma-positivity}. The operators $J(\gamma)$, $L_{c,p}(\gamma)$, and
$K_{c,p,b}(\gamma)$  satisfy the Hypothesis~\ref{h:1} in Section 2.1, and $K_{c,p,b}(\gamma)$ is positive when $p\not=0$,
according to  Lemma~\ref{lemma-positivity}. Consequently,  $K_{c,p,b}(\gamma)$ is nonnegative with trivial kernel, and the result in  Corollary \ref{cor-stability} implies that the operator $J(\gamma)L_{c,p}(\gamma)$ has no unstable spectrum.
This proves Theorem~\ref{t:spectral}~(i). The proof of the second part of  Theorem \ref{t:spectral} has been discussed in Section~3.1.

\begin{Remark}
  The abstract result in Corollary~\ref{cor-stability} allows to also recover the proof of spectral stability of the periodic traveling
wave $\phi_c$ as a solution of the KdV equation (\ref{kdv}). Indeed, for $p=0$, the operators $J(\gamma)$, $L_{c,0}(\gamma)$, and
$K_{c,0,b}(\gamma)$  satisfy the Hypothesis~\ref{h:1} in Section 2.1, and by Corollary \ref{cor-positivity}, $K_{c,0,b}(\gamma)$ is positive for $\gamma \in (0,1)$, and for $\gamma=0$ it is nonnegative and has a one-dimensional kernel spanned by~$\partial_x \phi_c$. Since $\partial_x \phi_c$ also belongs to the  kernel of $J(0) L_{c,0}(0)$, due to the translational invariance, the result in Corollary \ref{cor-stability} implies that
the operator $J(\gamma)L_{c,0}(\gamma)$ has no unstable spectrum, for any $\gamma \in [0,1)$. Consequently, the periodic traveling
wave $\phi_c$ is stable as a solution of the KdV equation (\ref{kdv}).
\end{Remark}

\subsection{Proof of Theorem~\ref{t:linear}}

Following the arguments in Sections~3.2, 4.2 and 4.3, we define the linear operator
\begin{equation}
\label{commuting-K}
K_c = M_c-bL_c,
\end{equation}
with
\[
M_{c} = \partial_x^4 + 10 \partial_x \phi_c(x) \partial_x - 10 c \phi_c(x) - c^2 - \frac{5}{3}\left(1 + c \partial_x^{-2} \right) \partial_y^2 ,
\]
$L_c$ given by \eqref{JLc},
and  some $b\in(b_0(c),b_+(c))$, as in Lemma~\ref{lemma-positivity}.
Then $K_c$ satisfies the commutativity property \eqref{e:lckc} with $J = \partial_x$,
and we claim that its restriction to the space $\dot L^2(N,p)$ is a nonnegative operator with one-dimensional kernel
spanned by the translation mode  $\partial_x\phi_c$. (Here again, the restriction to the  space $\dot L^2(N,p)$
means that $K_c$ as defined above is composed with the standard projection on the subspace of functions with zero mean.)
Indeed, it is not difficult to check that $K_c$ is a self-adjoint operator and using Fourier series in $y$,
and Floquet decomposition in $x$, that its spectrum is given by
\[
\sigma(K_c) = \bigcup_{n\in\Z}\bigcup_{\gamma\in\mathcal I_N} \sigma\left(K_{c,pn,b}(\gamma) \right),\quad \mathcal I_N = \left\{\frac jN\;;\; j=0,\dots,N-1\right\},
\]
with $K_{c,pn,b}(\gamma)$ being the operators defined by \eqref{e:Kcp-gamma}. Then the result in  Lemma~\ref{lemma-positivity} proves the claim.
As a consequence, there exists a positive constant $c_{N,p}$, such that
\[
\langle K_cw,w\rangle\geq c_{N,p}\|w\|^2, \quad \forall\ w\in \dot H^{2,1}(N,p),
\quad \langle w,\partial_x\phi_c\rangle = 0,
\]
where $\langle\cdot,\cdot\rangle$ and $\|\cdot\|$ denote the scalar product and the norm, respectively, in $\dot L^2(N,p)$.
G\r arding's inequality further implies that
\begin{equation}\label{e:gard}
\langle K_cw,w\rangle\geq c_{N,p}\|w\|^2_{2,1}, \quad \forall\ w\in H^{2,1}(N,p),
\quad \langle w,\partial_x\phi_c\rangle = 0,
\end{equation}
with a possibly different constant  $c_{N,p}$.

For a solution $w\in C^1(\R,H^{2,1}(N,p))$ to the linearized equation \eqref{e:linKP}, the equality \eqref{e:kwt} holds. We set
\[
w(t) = a(t) \partial_x\phi_c + w_1(t),\quad
a(t) = \frac{\langle w(t),\partial_x\phi_c\rangle}{\|\partial_x\phi_c\|^2}, \quad
 \langle w_1(t),\partial_x\phi_c\rangle = 0.
\]
Inserting this decomposition
into \eqref{e:kwt}, using the inequality \eqref{e:gard}, and the fact that  $\partial_x\phi_c$ spans the kernel of $K_c$,
we find
\begin{eqnarray}
\nonumber
c_{N,p}\|w_1(t)\|^2_{2,1}&\leq& \langle K_cw_1(t),w_1(t)\rangle =
\langle K_cw(t),w(t)\rangle \\ &=&  \langle K_cw(0),w(0)\rangle \leq C_{N,p}\|w(0)\|^2_{2,1},
\label{bound-w-1}
\end{eqnarray}
where $C_{N,p}$ exists due to the boundedness of the quadratic form (\ref{e:kwt}) in the energy space $H^{2,1}(N,p)$. This proves the first inequality in \eqref{bound-theorem}.

Next,  by taking the scalar product of the linearized equation \eqref{e:linKP} with $\partial_x\phi_c$ we obtain that $a(t)$ satisfies the first order differential equation
\begin{equation}\label{e:442}
a'(t) \; \| \partial_x\phi_c \|^2 = \langle \mathcal{B}_c w_1(t), \partial_x \phi_c \rangle = -\langle w_1(t), L_c \partial_x^2 \phi_c \rangle.
\end{equation}
The inequality \eqref{bound-w-1} above, together with  the Cauchy-Schwarz inequality, implies that the last term in \eqref{e:442} is a bounded function. Consequently, $a'(t)$ is bounded, which proves the second inequality in \eqref{bound-theorem} and completes the proof of Theorem \ref{t:linear}.

\section{Discussion}

The general counting result in Section~2 allowed us to prove the transverse spectral and linear stability of periodic waves
for the KP-II equation  (\ref{KP}).
In this section, we address the question of their transverse nonlinear stability which remains open.

It is tempting to construct a higher-order energy functional associated with the
 linear operator $M_{c,p}$ given by (\ref{operator-M-equivalent}), which could
 then be used for a nonlinear stability proof, just as for the KdV and NLS equations \cite{Decon1,b1,gp}. Since the part $M_{\rm KdV}$
in $M_{c,p}$ is the Hessian operator for $R_c(u)$ in (\ref{energy-R}),
which is constructed from the higher-order energy functional $H(u)$ in (\ref{higher-order-energy-KdV}),
whereas the part $M_{\rm KP}$ in $M_{c,p}$ has constant coefficients,
a higher-order energy functional can be thought in the following form
\begin{equation}\label{KP-higher-order-energy}
\widetilde{F}(u) = \int\int \left[ u_{xx}^2 - 10 u u_x^2 + 5 u^4 + \frac{5}{3} u_y^2 - \frac{5c}{3} (\partial_x^{-1} u_{y})^2 \right] dx dy.
\end{equation}
However, the function $\widetilde{F}(u)$ has a speed parameter $c$ in front of the
last term, which is also the last term of the energy functional $\widetilde{E}(u)$
for the KP-II equation (\ref{KP}) given by
\begin{equation*}
\widetilde{E}(u) = \int \int \left[ u_x^2 - 2 u^3 - (\partial_x^{-1} u_{y})^2 \right] dx dy.
\end{equation*}
Since $\widetilde{E}(u)$ is constant in time and the speed $c$ is an independent parameter,
the quantity $\widetilde{F}(u)$ in (\ref{KP-higher-order-energy}) is not related
to a conserved quantity of the KP-II equation (\ref{KP}).
Therefore, the commuting operator $K_c$ in (\ref{commuting-K}) constructed in this paper
is not the Hessian operator for a higher-order conserved quantity of the KP-II equation (\ref{KP}).

On the other hand,  for the KP-I equation, a conserved higher-order energy functional
has been constructed in \cite{Z1,Z2}. After transforming this quantity
to the variables used in the KP-II equation (\ref{KP}), it can be written in the form
\begin{equation*}
\widetilde{H}(u) = \int\int \left[ u_{xx}^2 - 10 u u_x^2 + 5 u^4 - \frac{10}{3} u_y^2 + \frac{5}{9} (\partial_x^{-2} u_{yy})^2
+ \frac{10}{3} u^2 \partial_x^{-2} u_{yy} + \frac{10}{3} u \left( \partial_x^{-1} u_y \right)^2 \right] dx dy.
\end{equation*}
Similarly to $\widetilde{F}(u)$, the $y$-independent part of $\widetilde{H}(u)$ is equivalent to
the higher-order energy functional $H(u)$ of the KdV equation (\ref{kdv}) given by (\ref{higher-order-energy-KdV}).
However, unlike $\widetilde{F}(u)$, the quantity $\widetilde{H}(u)$ is constant in time.

The periodic traveling wave $\phi_c$ is a critical point of the
higher-order energy functional $\widetilde{R}_c(u) = \widetilde{H}(u) - c^2 \widetilde{Q}(u) + 2 I \widetilde{C}(u)$,
where $\widetilde{Q}(u)$ and $\widetilde{C}(u)$ generalize $Q(u)$ and $C(u)$ by including the double integration in
$x$ and $y$. After a Fourier transform in the variable $y$, we find that
the Hessian operator at the periodic wave $\phi_c$ related to $\widetilde{R}_c(u)$ is given by
\begin{eqnarray}
\nonumber
\widetilde{M}_{c,p} & = & \partial_x^4 + 10 \partial_x \phi_c(x) \partial_x - 10 c \phi_c(x) - c^2 \\
& \phantom{t}& - \frac{10}{3} p^2 \left(1 + \phi_c(x) \partial_x^{-2} + \partial_x^{-1} \phi_c(x) \partial_x^{-1}
+ \partial_x^{-2} \phi_c(x) \right) + \frac{5}{9} p^4 \partial_x^{-4}.  \label{operator-M}
\end{eqnarray}
A long, but straightforward, symbolic computation shows that the commutativity condition (\ref{commutability}) is indeed satisfied with the
two linear operators $L_{c,p}$ and $\widetilde{M}_{c,p}$ given by (\ref{operator-L}) and (\ref{operator-M}), respectively.

Note that the expression (\ref{operator-M}) for the operator $\widetilde{M}_{c,p}$ is
different from the expression (\ref{operator-M-equivalent}) for the operator $M_{c,p}$
obtained by our direct symbolic computations. Clearly, the difference between these two operators,
\begin{eqnarray*} 
M_{c,p} -\widetilde M_{c,p} = \frac{5}{3} p^2 \left(3 + c \partial_x^{-2} + 2 \phi_c(x) \partial_x^{-2} + 2 \partial_x^{-1} \phi_c(x) \partial_x^{-1}
+ 2 \partial_x^{-2} \phi_c(x) \right) - \frac{5}{9} p^4 \partial_x^{-4},
\end{eqnarray*}
also satisfies the commutativity condition (\ref{commutability}). The operator equation (\ref{commutability}) admits multiple solutions, but the most general form for a solution $M_{c,p}$ is unknown.

In contrast to the operator $M_{c,p}$ given by (\ref{operator-M-equivalent}), the operator $\widetilde{M}_{c,p}$ in (\ref{operator-M}) cannot be used to construct  commuting positive operators, like the operators $K_{c,p,b}(\gamma)$ obtained in Section 4.3. Indeed, by using the Floquet--Bloch transform and by
taking a linear combination of the two operators $L_{c,p}(\gamma)$ and
$\widetilde{M}_{c,p}(\gamma)$ in the form
\begin{equation}
\widetilde{K}_{c,p,b}(\gamma) = \widetilde{M}_{c,p}(\gamma) - b L_{c,p}(\gamma),
\end{equation}
where $\gamma \in [0,1)$, we can check the analogue of property (\ref{operator-M-positivity}) in Remark \ref{rem-M-property}.
For $c = 1$, when $\phi_c = 0$, and $b = 2$, by using Fourier series in $x$,
we obtain the spectrum of $\widetilde{K}_{1,p,2}(\gamma)$,
\begin{equation}\label{e:57}
\sigma({\widetilde{K}}_{1,p,2}(\gamma)) = \left\{ \left( k^2 - 1 \right)^2
+ \frac{p^2 (5 p^2 - 30 k^4 + 18 k^2)}{9 k^4} \;;\; k=\gamma + n,\  n\in\Z,\ \gamma + n\not=0 \right\}.
\end{equation}
If $p = 0$, which corresponds to the KdV case, the operator ${\widetilde{K}}_{1,0,2}(\gamma)$ is nonnegative, for every $\gamma \in [0,1)$.
On the other hand, by inspecting the sign of the function in \eqref{e:57}, we can show that, for every $p \neq 0$, the operator ${\widetilde{K}}_{1,p,2}(\gamma)$ has some negative eigenvalues, at least for some values $\gamma \in [0,1)$, and then conclude that $\widetilde{K}_{c,p,b}(\gamma)$ is not always positive.

Summarizing, the existence of a Lyapunov functional for the KP-II equation (\ref{KP}) which could be used for a transverse nonlinear stability proof for periodic waves is not known, and this nonlinear stability problem remains open.
We point out that the analytical difficulty of using the higher-order energy functional $\widetilde{H}(u)$ for a nonlinear stability proof seems to be the same as the one arising in the proof of global well-posedness of the KP-II equation in the energy space (see \cite{Koch} and the references therein).

\end{document}